\documentclass[twocolumn,aps,JPCM,showpacs,superscriptaddress,unsortedaddress]{revtex4}
\usepackage{graphicx}
\usepackage{epsf}
\newcommand{\etal}{{\it et al.}}

\begin{document}

\title{The electronic structure of La$_{1-x}$Sr$_{x}$MnO$_{3}$ thin films
and its $T_c$ dependence as studied by angle-resolved photoemission}

\author{M. Shi}
\email{ming.shi@psi.ch} \affiliation{Swiss Light Source, Paul
Scherrer Institut, CH-5232 Villigen PSI, Switzerland}
\author{M. C. Falub}
\affiliation{Swiss Light Source, Paul Scherrer Institut, CH-5232
Villigen PSI, Switzerland}
\author{P. R. Willmott}
\affiliation{Swiss Light Source, Paul Scherrer Institut, CH-5232
Villigen PSI, Switzerland}
\author{J. Krempasky}
\affiliation{Swiss Light Source, Paul Scherrer Institut, CH-5232
Villigen PSI, Switzerland}
\author{R. Herger}
\affiliation{Swiss Light Source, Paul Scherrer Institut, CH-5232
Villigen PSI, Switzerland}
\author{L. Patthey}
\affiliation{Swiss Light Source, Paul Scherrer Institut, CH-5232
Villigen PSI, Switzerland}
\author{K. Hricovini}
\affiliation{Universit\'e de Cergy-Pontoise, 95031 cergy-Pontoise
CEDEX, France}
\author{C. V. Falub}
\affiliation{EMPA, Switzerland}
\author{M. Schneider} \affiliation{Laboratory for Neutron
Scattering, ETH Zurich and Paul Scherrer Institut, CH-5232 Villigen
PSI, Switzerland}

\begin{abstract}
We present angle-resolved photoemission spectroscopy results for
thin films of the three-dimensional manganese perovskite
La$_{1-x}$Sr$_{x}$MnO$_{3}$. We show that the transition temperature
($T_c$) from the paramagnetic insulating to ferromagnetic metallic
state is closely related to details of the electronic structure,
particularly to the spectral weight at the ${\bf k}$-point, where
the sharpest step at the Fermi level was observed. We found that
this ${\bf k}$-point is the same for all the samples, despite their
different $T_c$. The change of $T_c$ is discussed in terms of
kinetic energy optimization. Our ARPES results suggest that the
change of the electronic structure for the samples having different
transition temperatures is different from the rigid band shift.
\end{abstract}

\pacs{}
\date{\today}
\maketitle

Colossal magnetoresistance (CMR) in hole-doped manganese oxides with
perovskite structures~\cite{Kusters,Jin} is a phenomenon of great
scientific and technological importance. For a certain range of
doping, La$_{1-x}$Sr$_{x}$MnO$_{3}$ (LSMO) shows a large decrease in
resistivity upon cooling, associated with a paramagnetic (PM) to
ferromagnetic (FM) transition~\cite{Urushibara,Hemberger}. Close to
the transition temperature $T_c$, the resistivity can be further
strongly reduced by applying a magnetic field, in a phenomenon known
as colossal magnetoresistance. The temperature-dependent resistivity
in the FM phase have been qualitatively explained by the
double-exchange (DE) mechanism~\cite{Zener,Cieplak}. The premise is
as follows: in the FM phase, LSMO contains mixed-valent Mn$^{3+}$
and Mn$^{4+}$. For the site-symmetry of the cation in the MnO$_6$
octahedra, the valence states in question are Mn$^{4+}$:
$t_{2g}^{3}$ and Mn$^{3+}$: $t_{2g}^{3}$ $e_{g}^{1}$. There are
$(1-x)$ $e_{g}$ electrons per unit cell, which are free to move
through the crystal, subject to a strong Hund's coupling to the
localized Mn$^{4+}$ (S = 3/2) spins. The kinetic (band) energy is
minimized by making all the spins parallel. It was also realized
that the DE alone is not enough to explain CMR and other effects,
especially the insulator-like transition above $T_c$. The DE
framework should be supplemented with more refined ideas (e.g.
Jhan-Teller distortions, polaron formations)~\cite{Millis2}. Among
the Ruddelson-Popper series of manganites,
(La,Sr)$_{n+1}$Mn$_n$O$_{3n+1}$ (n = 1, 2, ...), (La,Sr)MnO$_3$ has
the highest $T_c$ and its resistivity at low temperatures is about
two orders of magnitude lower than that of the layered manganite (n
= 2)~\cite{Kimura}.

One major obstacle in understanding the physics of the
three-dimensional manganites has been a lack of detailed knowledge
of the electronic structure of the low binding energy electronic
states. So far, only very limited experimental results on ${\bf
k}$-resolved electronic structures of La$_{1-x}$Sr$_{x}$MnO$_{3}$
have been reported~\cite{Shi,Falub,Chikamatsu}. In this work we
apply angle resolved photoemission spectroscopy (ARPES) to probe the
electronic structure of metallic single crystalline films of
La$_{1-x}$Sr$_{x}$MnO$_{3}$ with different $T_c$. We observe that
the finite spectral weight at the Fermi level (E$_F$) is closely
associated with a broad peak which disperses at higher binding
energies. The difference in $T_c$ is directly reflected by a change
in the electronic structure of the lowest binding energy states.

\begin{figure*}
\includegraphics
[width=6.0in]
 {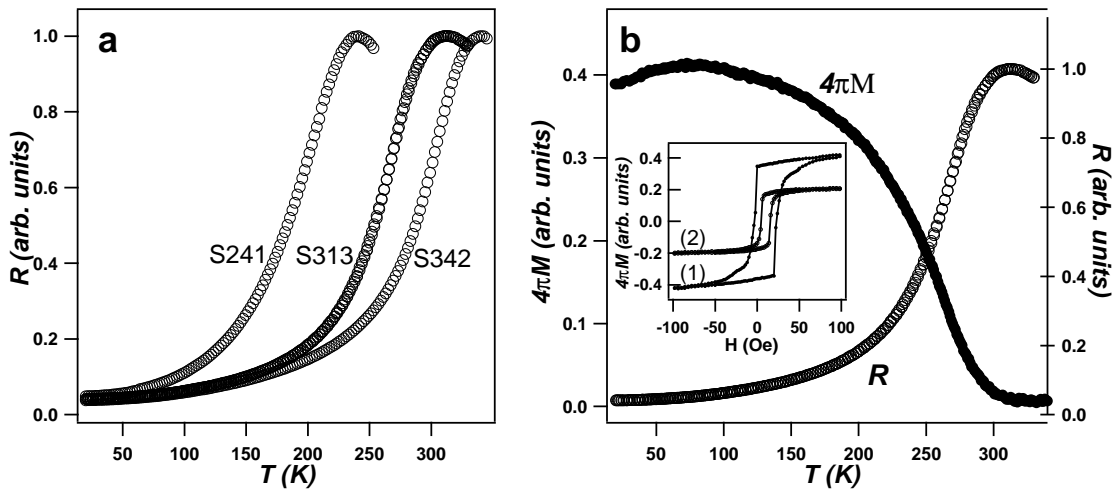}
\caption{a) Resistance vs. temperature for sample S241, S342 and
S313. The resistance was normalized to the peak value. b) Resistance
and DC magnetization of S313 vs. temperature. Inset: magnetic
hysteresis curves of S313 measured at temperatures 100 K (1) and 250
K (2).} \label{fig1}
\end{figure*}

1300 \AA-thick single crystalline thin films of LSMO  were prepared
by {\it in situ} heteroepitaxial growth on SrTiO$_3$ (001)
substrates by a novel adaptation of pulsed laser
deposition~\cite{Willmott1,Willmott2}. {\it In situ} reflection
high-energy electron-diffraction patterns and Kiessig fringes in
{\it ex situ} x-ray reflectivity curves demonstrate that the films
have a surface roughness of less than one monolayer. Low-energy
electron-diffraction analysis shows a clear (1 $\times$ 1) pattern
with no sign of surface reconstruction. Three LSMO samples with
ascending Sr/La ratios (0.44, 0.48, 0.52) were prepared. Because the
hole-doping level could be changed by the small variation of the
Oxygen stoichiometry, which can result in a change of $T_c$, we
further characterize the bulk properties of the samples by transport
measurements. Figure~1a shows the resistance-temperature curves
[R(T)] obtained from four-probe measurements. The transition
temperatures determined from these data are 241 K, 342 K and 313 K,
respectively. In the rest of the paper we will label the samples by
their $T_c$ as S241, S342 and S313, respectively. The transition
temperatures were confirmed by DC magnetization measurements.
Figure~1b shows the magnetic momentum [4$\pi$M(T)] of S313, together
with R(T), as a function of temperature. The stoichiometry of S313,
as determined by {\it ex situ} using Rutherford backscattering
spectrometry, was La$_{0.66}$Sr$_{0.34}$MnO$_{3}$.

\begin{figure}
\includegraphics
[width=\columnwidth]
 {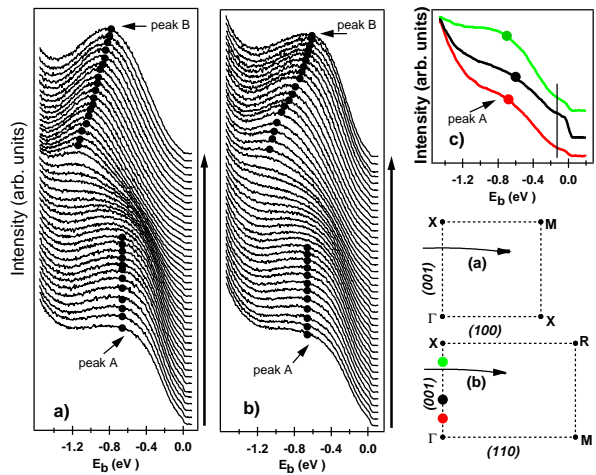}
\caption{ARPES spectra for S313 at 30 K.  (100) and (110) are along
the sample surface, while (001) is the surface normal. Peak A and
peak B are used to facilitate the discussion in the text. a) and b)
EDCs taken by using E$_{\it h \nu}$ = 44 eV with linearly
horizontally polarized light and with circularly polarized light,
respectively. The paths in the BZ are indicated with arrowed lines
in the middle and bottom-right of the figure. Circles indicate the
peak positions of broad peaks. c) EDCs taken with circularly
polarized light with E$_{\it h \nu}$ = 34 eV (red), 38 eV (black)
and 46 eV (green) in the normal emission. The ${\bf k}$-points are
indicated in the bottom-right of the figure with filled colour
circles. The vertical line indicates the energy where the slopes of
the EDCs change, circles indicate the peak positions of broad
peaks.} \label{fig2}
\end{figure}

ARPES measurements were performed at the Surface and Interface
Spectroscopy (SIS) beamline at the Swiss Light Source (SLS). During
measurements, the pressure always remained less than 1 $\times$
10$^{-10}$ mbar. The spectra were recorded with a Scienta 2002
analyzer with an angular resolution of better than 0.2$^\circ$. The
energy resolution was relaxed to 40 meV to obtain a high intensity.
All measurements were performed at temperature below 30 K. The
reduced zone scheme is used to indicate the wave vectors (${\bf k}$)
in reciprocal space. The free-electron final state approximation
with (V$_0$ - $\phi$ ) = 10.16 eV~\cite{Shi} is applied to determine
paths or ${\bf k}$-points, where V$_0$ is the inner potential and
$\phi$ is the work function. Photon energies (E$_{\it h \nu}$) are
indicated in the figure captions. In our ARPES measurements the
polar angles (between the surface normal and the direction of the
outgoing photoelectrons) were in the range of -5$^\circ$ $<$
$\theta$ $<$ 18$^\circ$.

\begin{figure}
\includegraphics
[width=\columnwidth]
 {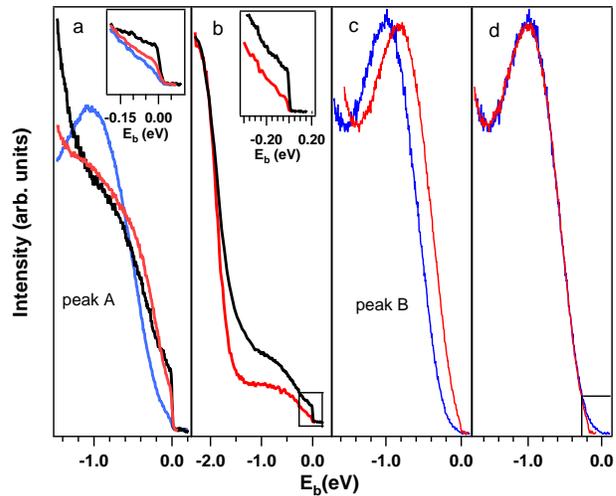}
\caption{EDCs for S241 (blue), S342 (black) and S313 (red). a) EDCs
at k $\approx$ (0,0,0.4) $\pi$/a taken with E$_{\it h \nu}$ = 38 eV,
normalized to the total area under EDCs. In inset spectra were
normalized to the intensity at E$_b$ = -200 meV. b) EDCs for S342
and S313 at {\bf k} =(0,0,0.4)$\pi$/a. The inset is a magnification
of the marked box. Spectra were normalized to the highest E$_b$
shown in the figure. c) EDCs for S241 (taken with E$_{\it h \nu}$ =
101 eV) and S313 (taken with E$_{\it h \nu}$ = 45 eV) at ${\bf k}$
$\approx$(0.6,0,0.7)$\pi$/a. d) the same as (c) with the EDC of S313
is offset by -170 meV. The labels of peak A and peak B follow the
peak assignments introduced in Fig. 2.} \label{fig3}
\end{figure}

We show representative ARPES spectra taken from S313 in Fig.~2.
Similar ARPES spectra were also obtained for S241 and S342. The
spectra were collected along paths parallel to the sample surface in
the (010) mirror plane (Fig.~2a) and (1$\bar{1}$0) mirror plane
(Fig.~2b), with linearly horizontally polarized light and with
circularly polarized light, respectively. The common features in
Fig.~2a and 2b are: close to the (001) axis a broad peak (peak A)
sits on the sloped background and shows nearly no dispersion along
k$_{||}$ (the component of ${\bf k}$ parallel to sample surface).
Away from the (001) axis at about k$_{||}$ = 0.3 $\pi$/a, another
peak (peak B) appears at higher binding energy and disperses towards
E$_F$ as k$_{||}$ increases. Further increasing k$_{||}$ does not
change the peak position of peak B in the (010) mirror plane, while
it folds back to high binding energy in (1$\bar{1}$0) mirror plane
(not shown). The bandwidth of peak B in the (1$\bar{1}$0) mirror
plane is larger than that in the (010) mirror plane. In contrast to
its non-dispersive behavior with respect to k$_{||}$ in the vicinity
of the (001) axis, peak A does show a dispersion with respect to
k$_\perp$ (along the surface normal, Fig.~2c). It can be seen that
there is a close correlation between the peak position and the
spectral weight at E$_F$. Specifically, when peak A approaches
E$_F$, the spectral weight at E$_F$ increases and the step at E$_F$
sharpens. It should be noted that a finite spectral weight at E$_F$
has only been observed in the vicinity of the (001) axis in the
Brillouin zone (BZ) with k$_{||}$ $<$ 0.4 $\pi$/a. The highest
spectral weight and the sharpest step at E$_F$ have been found at k
$\approx$ (0,0,0.4)$\pi$/a, when E$_{\it h \nu}$ = 38 eV or 67 eV is
used. This applies for all the investigated samples, despite their
different T$_c$. Peak B also disperses along a path parallel to the
(001) axis~\cite{Shi}. This is strong circumstantial evidence that
the dispersion of peak B derives from the bulk electronic structure.
However, the dispersive feature is much weaker than that in the
planes perpendicular to the (001) axis. It is important to mention
that we have traced the peak B in different mirror planes, as well
as many general ${\bf k}$-points in the BZ by using various photon
energies and photon polarizations. We found that there is a
correspondence between the positions of peak B and the $T_c$ of the
samples (Fig.~3c): the higher T$_c$ is, the closer becomes peak B to
E$_F$. However, for all three samples, the centroids of peak B never
approach closer than 0.6 eV to E$_F$, and there is never more than a
vanishingly small spectral weight at E$_F$.

Besides the many similarities, there are some quantitative
differences in the ARPES spectra of S241, S342 and S313. Figure~3a
and 3b show the EDCs taken with E$_{\it h \nu}$ = 38 eV,
corresponding to k $\approx$ (0,0,0.4)$\pi$/a, where the highest
spectral weight at E$_F$ was observed for all three samples. The
spectra were normalized to the total areas under the EDCs. An
important observation is that the spectral weight at E$_F$ is
closely related to the transition temperatures, namely, the spectral
weight at E$_F$ is higher when $T_c$ increases (Fig.~3a). In order
to remove any ambiguity when comparing the spectral weight at E$_F$,
two additional normalization methods were employed. First, to
minimize the contribution of the sloped background, we normalized
the EDCs to the intensity at E$_B$ = 200 meV below E$_F$ (the inset
of Fig.~3a). Second, for normalization we used the "shoulder" of the
Mn $t_{2g}$ states of Mn (Fig.~3b). In both cases the spectral
weights at E$_F$ for different samples have the same trend as the
$T_c$s of the samples. Figure~3a also shows that the line shape of
peak A changes dramatically for the samples with different $T_c$.
For S241, it has the lowest $T_c$ among the three samples, and peak
A is well defined. As $T_c$ increases some spectral weight is
transferred from high binding energies to that close to E$_F$, and
at the same time peak A becomes broader and less pronounced. On the
other hand, the line shape of peak B is rigid with respect to the
change of $T_c$. Figure~3c shows the EDCs for S241 and S313 taken at
the same k $\approx$(0.6,0,0.7)$\pi$/a where a single $e_g$ band is
expected~\cite{Livesay,Pickett}. The peak position of the EDC of
S313 is shifted about 170 meV towards E$_F$ with respect to that of
S241. After offsetting the EDC of S313 by -170 meV the two EDCs
overlap over nearly the entire energy range, except at the very low
binding energy tail, where the EDC of S313 has a smaller spectral
weight (the marked box in Fig.~3d).

We now address the issue of coherent electronic excitations at low
binding energies. In recent ARPES studies of the layered manganite
La$_{2-2x}$Sr$_{1+2x}$Mn$_{2}$O$_{7}$ with $x \approx 0.4$, it has
been shown that quasiparticle peaks do exist close to the Fermi
level~\cite{Sun,Mannella,Jong}. The quasiparticle peaks are followed
by incoherent excitations, and the single particle spectral function
has a "peak-dip-hump" structure which indicates strong many-body
effects and/or electron-lattice coupling. As our samples are
three-dimensional, one would expect a higher spectral weight of
coherent electronic excitations in the energy range where
quasiparticles were observed in the layered manganite. The absence
of sharp excitations in our ARPES spectra may be explained by the
uncertainty of k$_\perp$, the component of {\bf k} along the surface
normal, in the photoemission process. Because k$_\perp$ is not
conserved, due to the breaking of the translational symmetry at the
sample surface, ARPES spectra in a three-dimensional system will be
broadened ($\Delta$k$_\perp$), especially for an energy band that
has a large dispersion in the direction along the surface normal.
The large change of the slopes as marked by the vertical line in
Fig.~2c may indicate that the peak-dip-hump structure also exists in
our samples. However, the broadening effect smears out the dip in
the ARPES spectra.

We observed the finite spectral weight at E$_F$ only in the vicinity
of (0,0,0.4)$\pi$/a in the ${\bf k}$-space. This particular ${\bf
k}$-point would correspond to the ${\bf k}_F$ of the electron-pocket
centered at $\Gamma$ point from band structure
calculations~\cite{Livesay,Pickett} when assuming that
La$_{1-x}$Sr$_{x}$MnO$_{3}$ has a cubic structure. However, in our
measurements there is never more than a vanishingly small spectral
weight at E$_F$ at equivalent ${\bf k}$-points (0.4,0,0)$\pi$/a or
(0,0.4,0)$\pi$/a, despite that various photon polarizations and
different photon energies corresponding to the same ${\bf k}$-points
in the reduced zone scheme were used in our experiments. Finite
spectral intensities at the predicted Fermi surface of the
hole-pocket centered at the R point (the corner of the BZ) in those
calculations was also not observed in our measurements, and indeed
was not reported in other ARPES studies on the same
system~\cite{Chikamatsu}. As the calculated Fermi surface of the
hole-pocket has a cubic-like shape~\cite{Livesay,Pickett}, the
nesting of the Fermi surface between the parallel faces of the cube
may introduce an instability and result in a gap opening in the
electronic excitation spectra. The consequence of this will be that
the spectral weight at E$_F$ diminishes and the relevant energy band
folds back from E$_F$. In our previous ARPES studies on LSMO we
observed the folding back of an energy band which is supposed to
cross the calculated Fermi surface of the hole-pocket (see Fig.~1a
and Fig.~2a in ref.~\cite{Falub}). Another possibility to explain
the absence of the hole-pocket is heteroepitaxial strain in the
samples. The common point in our ARPES studies and others is that
single crystalline LSMO films grown on SrTiO$_3$ substrates were
used in the experiments. The tensile stress due to the lattice
mismatch between the LSMO and the substrate results in the ratio
between the lattice constant in the [001] direction and those in the
equivalent [100] directions is less than unity. Further studies are
needed to understand the discrepancy between the experimental and
calculated results, namely, is it due to the nesting, the change of
lattice constant, a combination of these, or other effects? To
clarify the role played by the strain induced by any lattice
mismatch, it will be very useful to perform ARPES measurements on
single crystals grown on different substrates, e.g. NdGaO$_3$. In
this case the compressive stress results in the lattice constant in
[001] direction being larger than that in [100] direction.
Comparison of the ARPES results on LSMO crystals under different
stress will be very important in understanding how the electronic
structure responds to small changes of the lattice constants.

Our ARPES results also indicate that the change of the electronic
structure for the samples with different $T_c$ is different than the
rigid band shift, as suggested in the angle integrated photoemission
on the LSMO films\cite{Horiba}. This is manifested by the facts that
a) for all the samples, the highest spectral weight was found at the
same ${\bf k}$-point (0,0,0.4)$\pi$/a when E$_{\it h \nu}$ = 38 eV
is used in the measurements; b) the centroid of peak A of S313 is
closer to E$_F$ than that of S342 at this ${\bf k}$-point, but S313
has a lower spectral weight at E$_F$ than S342 (Fig.~3a), which is
opposite to what one would expect from the rigid band
picture~\cite{Shen}; and c) for S241 the peak position of peak A at
{\bf k} $\approx$ (0,0,0.4)$\pi$/a is about the same as that of peak
B at {\bf k} $\approx$(0.6,0,0.7)$\pi$/a (Fig.~3a and 3c). When
going to S313, the shift of the peak position of peak B (170 meV) is
much smaller than that of peak A, $\sim$500 meV, and compared to the
small change in the line-shape of the peak B at the low
binding-energy tail (the marked box in the Fig.~3d) the spectral
weight of peak A undergoes a large redistribution (Fig.~3a).

The relationship between $T_c$ and the spectral weight at E$_F$,
namely that a higher $T_c$ is associated with more spectral weight
at E$_F$, can be qualitatively explained by the double exchange
model~\cite{Kubo,Ohata,Millis}. The essential quantity for the
transition from PM to FM is the metallic density of charge carriers,
which are subject to Hund's rule and other interactions. The scale
of the transition temperature is set by the kinetic energy of the
mobile charge carriers, which is proportional to the expectation
value of the hopping Hamiltonian. The highest $T_c$ is obtained when
the kinetic energy is maximized in the system. As the Fermi level
divides the occupied states and unoccupied states of electrons, the
spectral weight in electronic excitation spectra is directly related
to the hopping probability given by the number of electrons that are
free to move from site to site and the number of available empty
states that the electrons can hop into.

In summary, our ARPES measurements on LSMO thin films with different
$T_c$ reveal both common features and quantitative differences in
their electronic structures in the FM phase. It was found that $T_c$
is closely related to the spectral weight at E$_F$ at ${\bf k}$
$\approx$ (0,0,0.4) $\pi$/a, where the sharpest step at E$_F$ was
observed for all investigated samples. We also provide evidence that
the rigid band picture cannot account for the change of the
electronic structure for the samples having different $T_c$
resulting from the change of dopings.

This work was performed at the Swiss Light Source, Paul Scherrer
Institut, Villigen, Switzerland. R. Betemps, M. Kropf, F. Dubi and
J. Rothe are acknowledged for technical support. This work was
supported by Paul Scherrer Institut.

\end{document}